\documentclass{aa}
\pdfoutput=1
\usepackage{amsmath}
\usepackage{txfonts}
\usepackage[utf8]{inputenc}
\usepackage{natbib}
\bibpunct{(}{)}{;}{a}{}{,}
\usepackage[french,english]{babel}
\usepackage[breaklinks]{hyperref}
\usepackage{graphicx}
\graphicspath{{fig/}}

\newcommand*{\prn}{\ensuremath{\mathrm{Pm}}}
\newcommand*{\ren}{\ensuremath{\mathrm{Re}}}
\newcommand*{\rmn}{\ensuremath{\mathrm{Rm}}}

\begin{document}

\title{Intermittent turbulent dynamo \\ at very low and high magnetic Prandtl numbers}
\titlerunning{Intermittent turbulent dynamo at low and high magnetic Prandtl numbers}
\author{Éric Buchlin\inst{1,2}}

\institute{CNRS, Institut d'Astrophysique Spatiale, UMR8617, 91405 Orsay, France \email{eric.buchlin@ias.u-psud.fr} \and Univ Paris Sud, Institut d'Astrophysique Spatiale, UMR8617, 91405 Orsay, France}

\date{Received date / Accepted date}

\abstract{ 
  Direct numerical simulations of plasmas have shown that the dynamo effect is efficient even at low Prandtl numbers, i.e., the critical magnetic Reynolds number $\rmn_c$ that is necessary for a dynamo to be efficient becomes smaller than the hydrodynamic Reynolds number \ren\ when $\ren \rightarrow \infty$.
}{ 
  We test the conjecture that $\rmn_c$ tends to a finite value when $\ren \rightarrow \infty$, and we study the behavior of the dynamo growth factor $\gamma$ at very low and high magnetic Prandtl numbers.
}{
  We use local and nonlocal shell models of magnetohydrodynamic (MHD) turbulence with parameters covering a much wider range of Reynolds numbers than direct numerical simulations, that is of astrophysical relevance.
}{ 
  We confirm that $\rmn_c$ tends to a finite value when $\ren \rightarrow \infty$. As $\rmn \rightarrow \infty$, the limit to the dynamo growth factor $\gamma$ in the kinematic regime follows $\ren^\beta$, and, similarly, the limit for $\ren \rightarrow \infty$ of $\gamma$ behaves like $\rmn^{\beta'}$, with $\beta\approx\beta'\approx 0.4$.
}{ 
  Our comparison with a phenomenology based on an intermittent small-scale turbulent dynamo, together with the differences between the growth rates in the different local and nonlocal models, indicate that nonlocal terms contribute weakly to the dynamo effect.
}
  

\maketitle

\section{Introduction}
\label{sec:introduction}

The magnetic Prandtl number $\prn$ (the ratio of the kinematic viscosity $\nu$ to the magnetic diffusivity $\eta$) is one of the non-dimensional parameters that control the properties of plasma. This ratio translates to $\prn = \rmn / \ren$, where $\ren$ and $\rmn$ are the hydrodynamic and the magnetic Reynolds numbers respectively, and this means in particular that it is an important parameter influencing the properties of turbulence in a plasma. Its value is large in hot, rarefied plasmas such as the interstellar or intracluster medium and the solar wind, and it is small in cool and dense plasmas as in planets and the Sun's convective zone, which is at the origin of the solar magnetic field.

\citet{iskakov07} and \citet{scheko07} demonstrated with direct numerical simulations that the turbulent dynamo is effective even for small $\prn$, but the minimum Prandtl number that they were able to use was limited by the resolution of the simulation to $0.070$ (with eighth-order hyperviscosity).  Owing to this limitation, they cannot tell for sure whether the critical magnetic Reynolds number $\rmn_c$ tends to a finite value when the hydrodynamic Reynolds number \ren\ tends to infinity (their maximum \ren\ is 6200). Large-scale dynamos at \prn\ down to $10^{-3}$ have also been obtained by \citet{brandenburg09a}. However, the range of \prn\ that is relevant to astrophysics is even wider.

shell models of magnetohydrodynamical (MHD) turbulence allow us to go much beyond these limits, up to $\ren \approx 10^{12}$ and $\prn \approx 10^{\pm 12}$ in this paper.  They are dynamical models of the nonlinear interactions between fields on different scales and have been developed for the study of turbulence in various frameworks such as hydrodynamics \citep{gle73}, MHD \citep{glo85,yam87}, reduced MHD \citep{nig04,buchlin07a}, Hall-MHD \citep{galt07}, and MHD with a global rotation rate \citep{perrone11a}.  A review of shell models can be found in \citet{biferale03} and in the book by \citet{bohr05}.

The simplification that they provide, making them complementary to direct numerical simulations (that are limited by the resolution one can afford), has allowed numerous results to be obtained.  MHD shell models have been proven to display dynamo action \citep{glo85,fri98,sahoo10a}, including magnetic field reversals \citep{perrone11a}. They have in particular been used by \citet{stepanov06} for the dynamo at low $\prn$, and later by the same authors \citep{stepanov08a} to propose a phenomenology of the turbulent dynamo at both low and high $\prn$ \citep[using nonlocal shell models:][]{plunian07}.

In this paper, we use local and nonlocal shell models to derive properties of the dynamo over a very wide range of Reynolds and Prandtl numbers, extending the results of \citet{iskakov07} and \citet{stepanov06}.

\section{Model equations and numerical set up}
\label{sec:numerical}

In shell models, the Fourier space for the fields of MHD is divided in concentric shells of radii $k_n= \lambda^n$ ($\lambda$ is the separation factor between shells), and the fields in each of these shells are represented by the complex scalars $u_n$ for the velocity field and $b_n = B_n / \sqrt{\mu_0 \rho}$ for the magnetic field ($\rho$ is the density of the plasma, $b_n$ has the dimension of a velocity). The nonlinear terms of incompressible MHD, a convolution in Fourier space, are written in the following symmetric form when expressed as a function of the Elsässer variables $z_n^\pm = u_n \pm b_n$
\begin{equation}
  \label{eq:shell}
  \left(\mathrm{d}_t z_n^\pm\right)_{\textrm{NL}} = \frac{i k_n}2 \big(
  Q_n(z^\pm,z^\mp,a-b) + Q_n(z^\mp,z^\pm,a+b)\big) .
\end{equation}
In general, the coefficients of the nonlinear terms are determined by
the conservation of quantities representing the invariants of
incompressible MHD in three dimensions (3D), namely the energy, cross
helicity, and magnetic helicity.

In the local ``GOY'' shell model \citep{giu98,stepanov06}, the nonlinear terms are limited to quasi-local interactions between three consecutive shells
\begin{equation}
  Q_n(X,Y,c) \equiv c_1 X_{n+1}^*Y_{n+2}^* + c_2 X_{n-1}^*Y_{n+1}^* + c_3
  X_{n-2}^* Y_{n-1}^*
\end{equation}
with $c = a \pm b$ and
\begin{align}
  a_1 = 1 \qquad a_2 &= (1-\lambda) / \lambda^2 \qquad a_3 = -1 /
  \lambda^2 \notag \\
  b_1 = b_2 &= b_3 = 1 / \lambda(\lambda+1) . \notag
\end{align}

In the nonlocal ``Sabra'' model \citep{plunian07,stepanov08a}, where
the hypothesis of quasi-local interactions is released,
\begin{align}
  Q_n(X,Y,c) &\equiv \sum_{m=1}^N T_m \big (c_m^1 X_{n+m}^*Y_{n+m+1} + c_m^2
  X_{n-m}^*Y_{n+1} \nonumber \\
  & \qquad\qquad + c_m^3 X_{n-m-1} Y_{n-1}\big)
\end{align}
with the coefficients
\begin{align}
  a_m^1 = \lambda^m (\lambda+1) \quad a_m^2 = -\lambda -
  (-\lambda&)^{-m} \quad  a_m^3 = (1 - (-\lambda)^{-m}) / \lambda \notag \\
  b_m^1 = (-1)^{m+1} \qquad b_m^2 &= 1 \qquad b_m^3 = -1  \notag \\ 
  T_m = k_{m-1}^\alpha &/ \lambda(\lambda+1)  \notag
\end{align}
and where $N$ is the number of shells used in the computation. The $\alpha$ exponent in $T_m$ controls the strength of long-range (in Fourier space) nonlinear interactions: following \citet{stepanov08a}, we use $\alpha=-1$ for strong nonlocal interactions, and $\alpha=-5/2$ for weak nonlocal interactions; $\alpha = -\infty$ would correspond to no nonlocal interactions, i.e., to a local model.

The forcing $f_n$ is a solution of a stochastic Langevin equation and is applied to the two first modes $n = 0$ and $1$ of the velocity only, and the dissipation is modeled by Laplacian diffusivity coefficients $\nu$ (viscosity) and $\eta$ (magnetic diffusivity).

The equations for all 3 models (local GOY model, weakly and strongly nonlocal Sabra models) are solved numerically using a 3rd-order Runge-Kutta scheme for the nonlinear and forcing terms, and a first-order implicit scheme for the dissipation.  The time step is adaptive and is set to the smallest time scale of the nonlinear terms, with a security factor of 5.  We use $\lambda=2$, and the initial condition for the velocity field is $u_n=|u_0| e^{i\phi_n} k_n^{-1/3} e^{-\nu k_n^2 \delta t_0}$, where $|u_0| = 10^{-1}$, $\delta t_0=10^{-3}$, and $\phi_n$ are independent random phases. After one large eddy turnover time $t_1$ with no magnetic field (the shell-model equations reduce to hydrodynamic shell-model equations), we introduce an initial magnetic field $b_n=|b_0| e^{i\phi'_n} k_n^{-1/3} e^{-\eta k_n^2 \delta t_0}$, where $|b_0| = 10^{-10}$ and $\phi'_n$ are also independent random phases: the cross helicity is close to zero, and the Lorentz force (scaling as $k^2ub$) is negligible. The initial conditions for both $u$ (at $t=0$) and $b$ (at $t=t_1$) correspond to power-law energy spectra with a slope $-5/3$ cut by the equivalent of the dissipation during the duration $\delta t_0$.

Each model is run with a wide range of parameters $\nu$ and $\eta$ from $10^{-12}$ to $1$ (6281 independent runs of each model).  Such a range of parameters is only made possible thanks to the simplifications operated in the shell models.

Every ten time steps, we compute the kinetic energy $E_u = \frac12 \sum_n |u_n|^2$, from which we evaluate the hydrodynamic and magnetic Reynolds numbers by $\ren = 2\pi\sqrt{2E_u} / k_0 \nu$ and $\rmn = 2\pi\sqrt{2E_u} / k_0\eta$.  We also evaluate the growth rate $\gamma$ of the magnetic energy $E_b = \frac12 \sum |b_n|^2$ from the local slope of $\ln E_b(t)$.  We stop this analysis when $E_b$ becomes greater than $10^{-4}E_u$, i.e., restricting ourselves to the kinematic regime of the dynamo.  In this way, we get a large set of $(\ren, \rmn, \gamma)$ triplets, from which statistics can be evaluated.

\section{Results}

\paragraph{Growth rate as a function of Reynold numbers.}

\begin{figure}[tp]
  \centering
  {\includegraphics[width=.97\linewidth,height=.2\linewidth]{%
      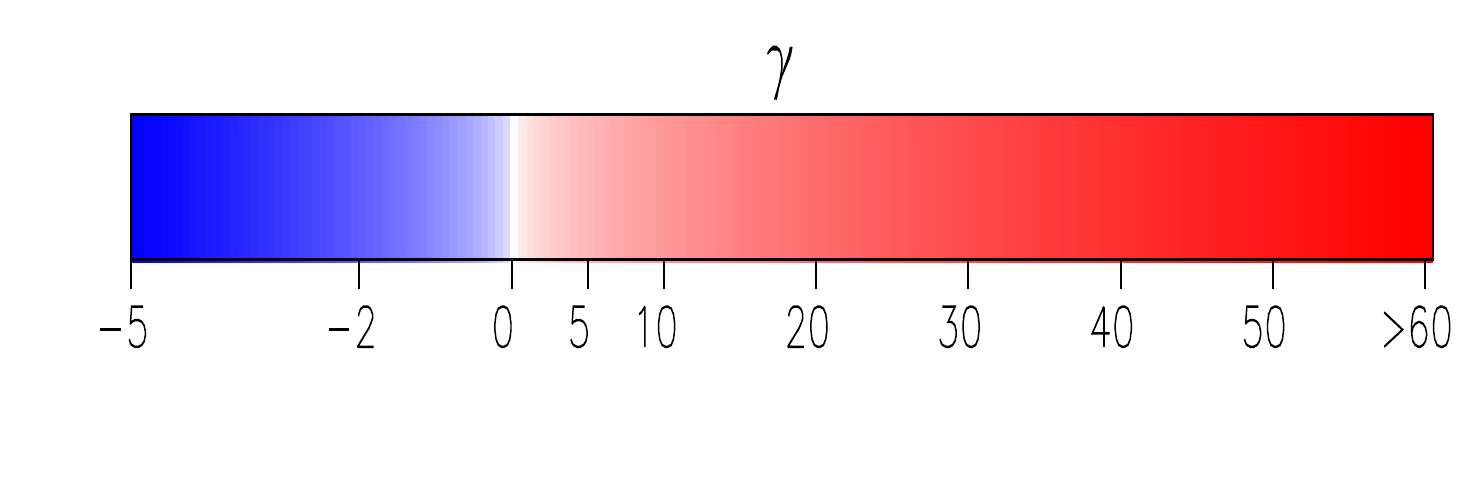}}\\[-.08\linewidth]
  {\includegraphics[width=\linewidth]{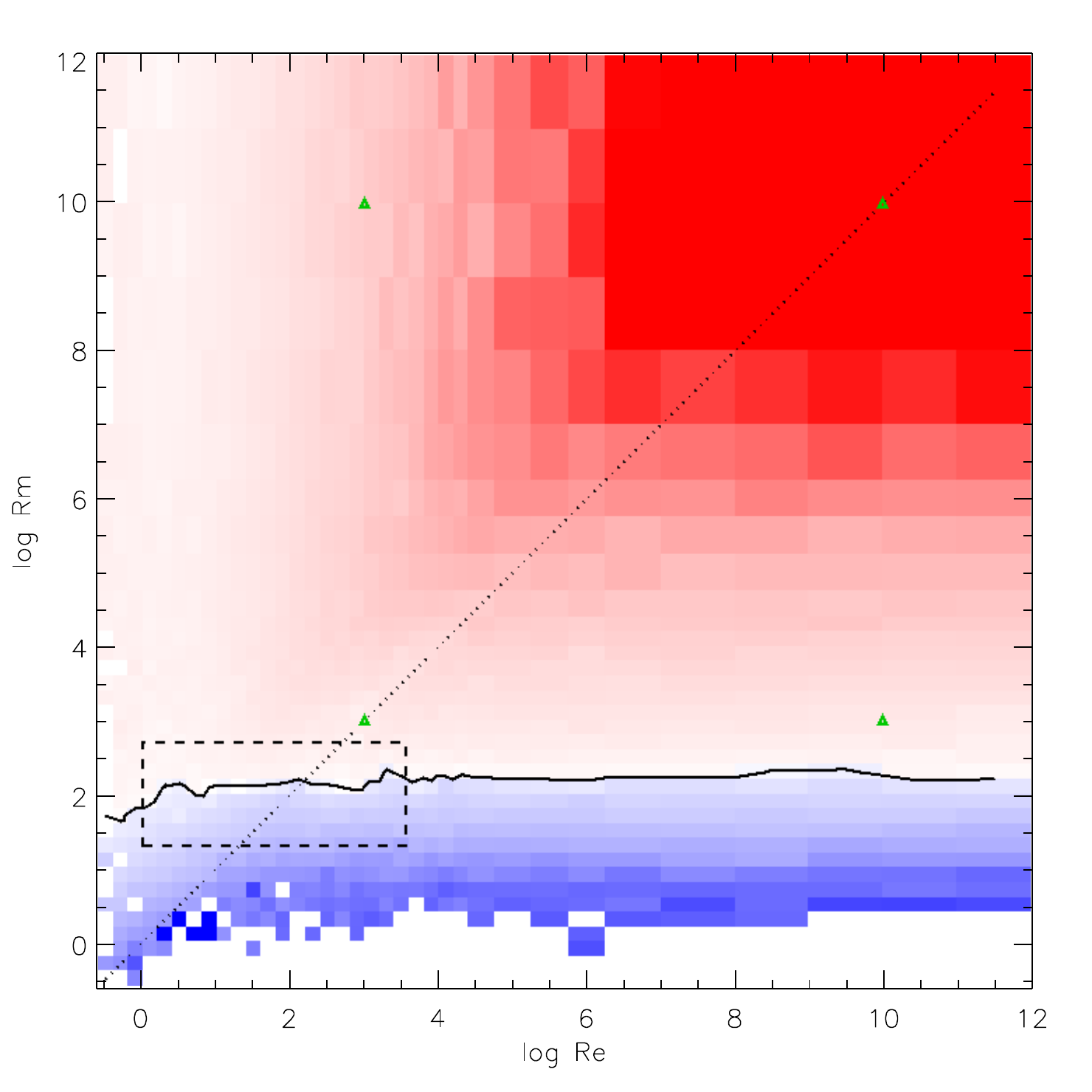}}
  \caption{Growth rate $\gamma$ as a function of the Reynolds numbers $\ren$ and $\rmn$ (note that the color scale is different for positive and negative values of $\gamma$), in the case of the weakly nonlocal Sabra model ($\alpha = -5/2$).  The stability curve, corresponding to the $\gamma=0$ level line, is overplotted.  The dashed rectangle corresponds to the range of parameters explored by the direct numerical simulations of \cite{iskakov07}; the green triangles indicate the initial Reynolds numbers used in Fig.~\ref{fig:spec}; the dotted line is where $\prn=1$.}
  \label{fig:gamma}
\end{figure}

In Fig.~\ref{fig:gamma}, we plot the average of $\gamma$ in bins of $(\ren,\rmn)$ for the weakly nonlocal model ($\alpha=-5/2$).  We see that the dynamo is effective ($\gamma>0$) for higher values of $\rmn$.  The level line $\gamma=0$ represents the stability curve $\rmn_c(\ren)$ of the dynamo; a striking feature is that it is almost independent of $\ren$ and that it seems to have a finite limit when $\ren \rightarrow \infty$.  This is a feature that has been noted by \citet{iskakov07} up to $\ren \approx 3600$, and here we confirm this result up to values of $\ren\approx 10^{12}$.  In addition, at low $\ren$, we also recover the decrease in the stability curve.

\begin{figure}[tp]
  \includegraphics[width=.92\linewidth]{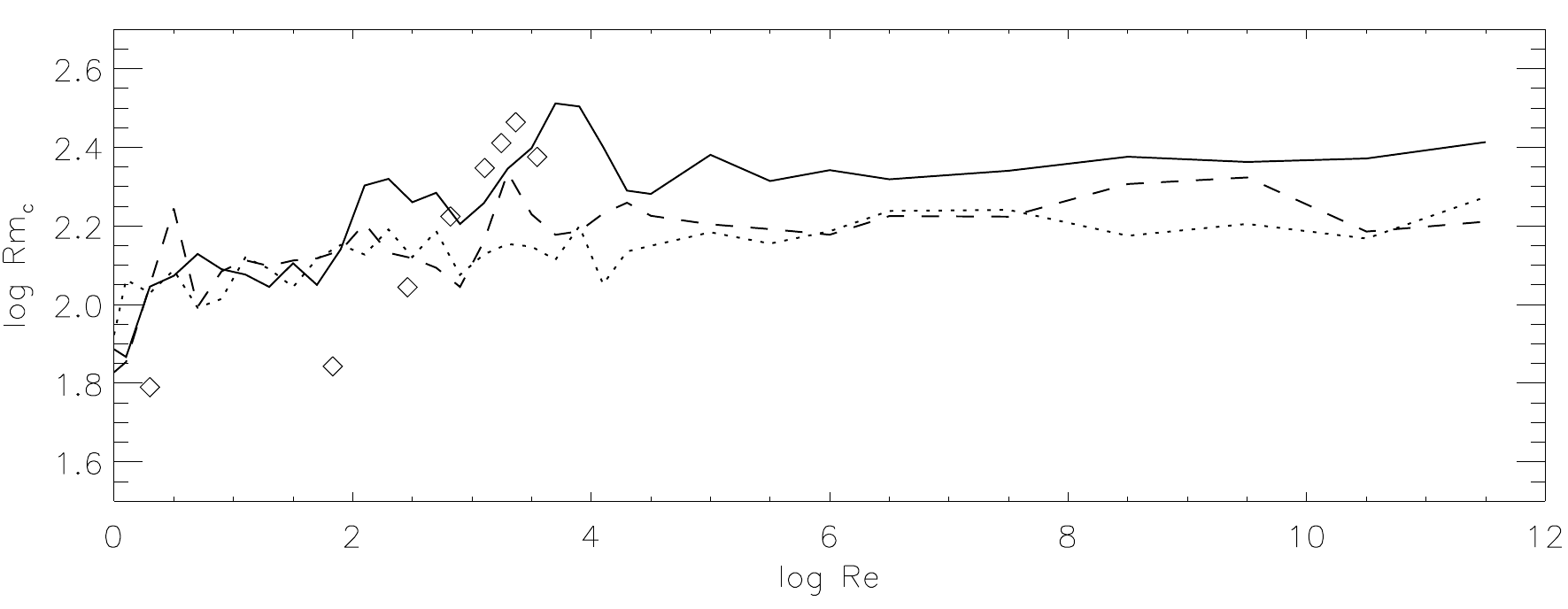}
  \caption{Stability curves $\rmn_c(\ren)$ for the local (plain line), weakly nonlocal (dashes), and strongly nonlocal (dots) models, and from \citet{iskakov07} (diamonds).}
  \label{fig:stab}
\end{figure}

The three different models (see also Fig.~\ref{fig:gamma2} and~\ref{fig:gamma3} online) have the same general behavior and the numerical results that we obtain for $\rmn_c$, summarized in Table~\ref{tab:values} are almost indistinguishable.  One can note however that the stability curves $\rmn_c (\ren)$ are shifted upwards (Fig.~\ref{fig:stab}) and that the limit of $\rmn_c$ as $\ren \rightarrow \infty$ slightly increases (Table~\ref{tab:values}) when the range of nonlocal interactions decreases. This implies that nonlocal interactions increase the efficiency of the dynamo, even at small $\prn$ where these nonlocal interactions are expected to be weak.

\begin{table}[tp]
  \centering

  \caption{Numerical values obtained for the three models:  $\ren_{\textrm{min}}$ is the smallest $\ren$ in Fig.~\ref{fig:gamma}, $\beta$ is the slope of the power-law fit to $l(\rmn) \equiv \lim_{\ren \rightarrow \infty}\gamma$, $\beta'$ is the slope of  the power-law fit to $l'(\ren) \equiv \lim_{\rmn \rightarrow \infty}\gamma$, $\zeta_1$ is the exponent of the first structure function for the velocity field, and $\beta_{\textrm{phenom.}}$ is the expected value for $\beta$ and $\beta'$ deduced from the small-scale dynamo phenomenology. Uncertainties in $\beta$, $\beta'$, and $\zeta_1$ are 3-$\sigma$ uncertainties in the fits corresponding to these parameters.} 

  \begin{tabular}{cccc}
    \hline \hline
    & Local GOY & Sabra, $\alpha=-5/2$ & Sabra, $\alpha=-1$ \\ 
    & Local & Weak nonloc. & Strong nonloc. \\\hline
    $\displaystyle{\lim_{\ren \rightarrow \infty} \log\rmn_c} $ &
    $2.4 \pm 0.2$ & $2.3 \pm 0.2$ & $2.2 \pm 0.2$ \\
    $\log\rmn_c(\ren_{\textrm{min}})$ & $1.5 \pm 0.3$ & $1.6 \pm 0.3$
    &  $1.6 \pm 0.3$  \\
    $\beta$   & $0.424 \pm 0.024$ & $0.418 \pm 0.032$ & $0.379 \pm 0.024$ \\
    $\beta'$  & $0.422 \pm 0.032$ & $0.379 \pm 0.032$ & $0.336 \pm 0.027$ \\
    $\zeta_1$ & $0.354 \pm 0.067$ & $0.350 \pm 0.067$ & $0.383 \pm 0.067$ \\
    $\beta_{\textrm{phenom.}}(\zeta_1)$ & $0.477 \pm 0.073$ & $0.481 \pm 0.073$ & $0.446 \pm 0.070$ \\
    \hline
  \end{tabular}

  \label{tab:values}
\end{table}

\paragraph{Growth rates at a given kinetic or magnetic Reynold number.}

\begin{figure}[tp]
  \centering
  \includegraphics[width=.92\linewidth]{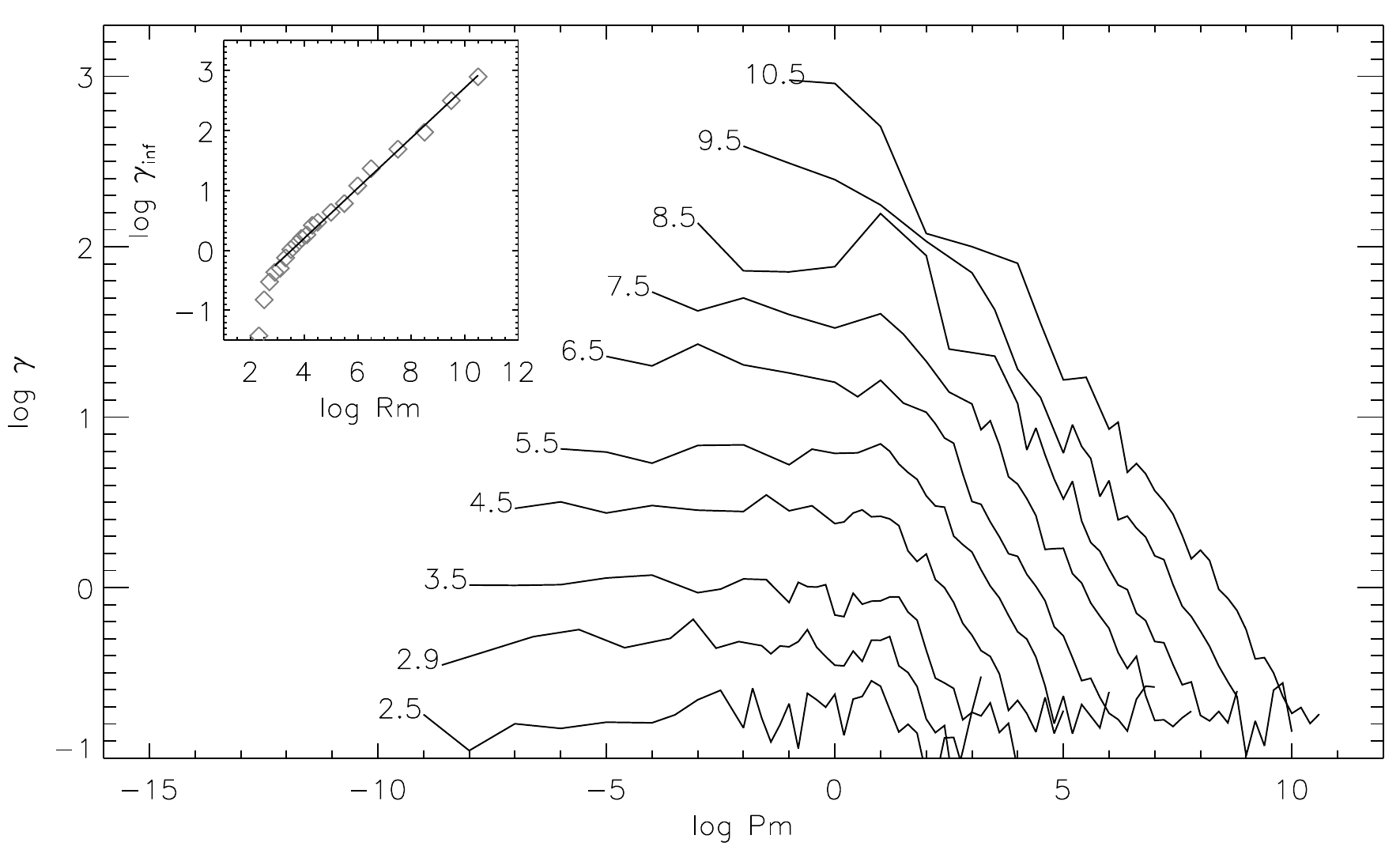}
  \caption{Logarithm of the growth rate $\gamma$ (when positive) as a function of the magnetic Prandtl number $\prn$ for different magnetic Reynolds numbers $\rmn$ (curves are labeled with $\log \rmn$), for the weakly nonlocal model. Inset: the limit $\gamma_\infty$ of $\gamma$ for $\ren \rightarrow \infty$ and a power-law fit (plain line).}
  \label{fig:gre}
\end{figure}

The curves $\gamma(\prn)$ for different $\rmn$ are plotted in Fig.~\ref{fig:gre}; the plot variables are the same as in Fig.~1 of \cite{iskakov07}.  These curves correspond to horizontal cuts in Fig.~\ref{fig:gamma}, with inverted abscissa and with a shift by $\log \rmn$ to the left.  They confirm two results of \cite{iskakov07} for the growth rate at a given \rmn, again over a much wider range of parameters: (1) for a given \rmn, the growth rate $\gamma$ reaches a limit\footnote{This limit is noted $\gamma_\infty (\rmn)$ in \cite{iskakov07}, but we need to use a different notation because of the other limit that we introduce.} $l(\rmn) \equiv \lim \gamma (\ren,\rmn)$ for $\ren \rightarrow \infty$ (i.e., $\prn \rightarrow 0$, the saturation of $\gamma$ starting for $\prn$ slightly greater than $1$); and (2) for any sufficiently large magnetic Reynolds number ($\rmn \gtrsim 250$), this limit $l(\rmn)$ is positive.

Similarly, vertical cuts in Fig.~\ref{fig:gamma} (not shown) demonstrate that for a given Reynolds number \ren, the growth rate $\gamma$ reaches a limit $l'(\ren) \equiv \lim \gamma (\ren, \rmn)$ for $\rmn \rightarrow \infty$, i.e., for $\prn \rightarrow \infty$. Again, the saturation of $\gamma$ starts for $\prn$ slightly greater than $1$; this asymmetry with respect to $\prn=1$ can be seen in Fig.~\ref{fig:gamma} as a shift between the dotted line ($\prn=1$) and the ``crest line'' of the representation of $\gamma (\ren,\rmn)$, and to our knowledge this shift has not been noted before.

\paragraph{Asymptotic behavior of the limits of the growth rate.}

We determine the limit $l(\rmn)$ as defined before by taking the average of $\gamma$ on the three leftmost data points of each curve of Fig.~\ref{fig:gre}, with the condition that a limit has been reached. The inset of Fig.~\ref{fig:gre} shows that $l(\rmn)$ is a power law of $\rmn$: its slope $\beta$ is obtained by a linear fit $\log l \sim \beta \log \rmn$, and the results for all three models are given in Table~\ref{tab:values}. Similarly, $l'(\ren)$ is a power law of $\ren$, and its slope $\beta'$ (also given in Table~\ref{tab:values}) is obtained by the linear fit $\log l' \sim \beta' \log \ren$.

Values of $\beta$ and $\beta'$ are in the range $[0.33,0.43]$. These values for $\beta$ and $\beta'$, which could not be determined by the direct numerical simulations of \citet{iskakov07}, seem to favor a dynamo driven by small-scale motions (an exponent $1/2$ is expected)
over an outer-scale, or mean-field, dynamo (an exponent of $0$ is
expected).  Assuming that $|u(\ell)| \sim \ell^{\zeta_1}$ (i.e.,
$\zeta_1$ is the exponent of the first structure function of the
velocity field), a small-scale dynamo phenomenology \citep[e.g.,][]{stepanov08a} can indeed be summarized as follows:
\begin{itemize}
  \item For $\prn \ll 1$, the scale on which the magnetic field grows the fastest is the resistive scale $\ell_\eta \sim \rmn^{-1/(1 + \zeta_1)}$, which lies within the inertial range of the velocity spectrum.  The growth rate is the inverse turnover time at this scale, i.e., $\gamma = u_{\ell_\eta} / \ell_\eta \sim \rmn^{(1 - \zeta_1) / (1 + \zeta_1)}$, giving $\beta = (1 - \zeta_1) / (1 + \zeta_1)$.
  \item Similarly, for $\prn \gg 1$, the scale on which the magnetic field grows the fastest is the viscous scale $\ell_\nu \sim \ren^{-1/(1 + \zeta_1)}$, where resistive dissipation is negligible. The growth rate is then $\gamma = u_{\ell_\nu} / \ell_\nu \sim \ren^{(1 - \zeta_1) / (1 + \zeta_1)}$, giving $\beta' = (1 - \zeta_1) / (1 + \zeta_1) = \beta$.
\end{itemize}

With no intermittency, $\zeta_1 = 1/3$ and we recover $\beta = \beta' = 1/2$ for the small-scale dynamo, as mentioned before. With hydrodynamic turbulence intermittency\footnote{A MHD model such as \citet{polit95} would not be relevant to the kinematic regime of the dynamo (low magnetic field).}, the \citet{shel94} phenomenology provides the value $\zeta_1 = 1 / 9 + 2 (1 - (2/3)^{1/3}) \approx 0.364$ and then the small-scale dynamo phenomenology gives $\beta = \beta' \approx 0.466$, in agreement with the numerical result of \citet{stepanov08a}.

In our simulations, the values we obtain for $\zeta_1$ (shown in Table \ref{tab:values} and computed from ten independent runs with $\nu=\eta=10^{-10}$) are closer to the \citealt{shel94} value than to one-third (no intermittency), although both are within 3-$\sigma$ error bars. The small-scale dynamo phenomenology then yields the values for $\beta$ and $\beta'$ given as $\beta_{\textrm{phenom.}}(\zeta_1)$ in Table \ref{tab:values}. This prediction and the simulation values are lower in the case of the strongly nonlocal model, as a consequence of the higher value of $\zeta_1$ in this case.

The values for $\beta$ and $\beta'$ are systematically lower than the $\beta_{\textrm{phenom.}}$ predictions, but they are still mostly consistent with them. They are also lower than the numerical result of \citet{stepanov08a}. The difference from the prediction might be caused, at least for the nonlocal models, by a contribution of the outer-scale dynamo to the growth of the magnetic field, while the difference from the numerical results of \citet{stepanov08a} may come from the different averaging process\footnote{At each time step, these authors start several simulations from the same initial fields, which are the average of the fields computed in at the previous time step, while we compute the growth rates from completely independent runs of the simulations and compute the average growth rate in the end.}.

\paragraph{Magnetic energy spectra.}

\begin{figure}[tp]
  \centering
  \includegraphics[width=.9\linewidth]{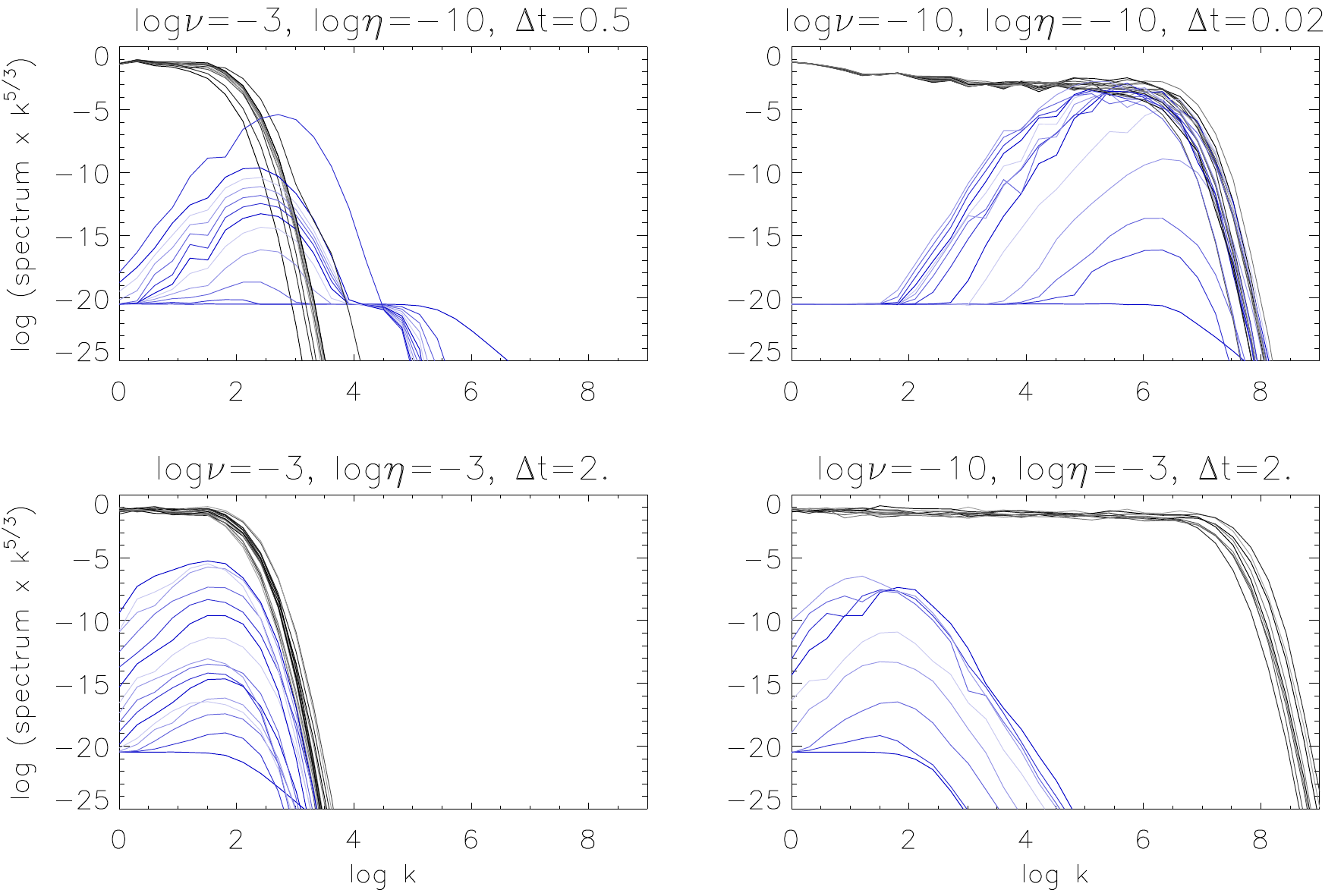}
  \caption{Evolution of the spectra (compensated by $k^{5/3}$) of kinetic energy (black) and magnetic energy (blue), for different sets of initial Reynolds numbers $(\ren,\rmn)$ (shown as green triangles in Fig.~\ref{fig:gamma}), for the weakly nonlocal Sabra model. The spectra are averaged over ten independent runs and are plotted at time intervals shown as $\Delta t$ in the plot titles. The different shades are a guide for understanding the direction of time (from black to gray and from dark to light blue, cycling every five spectra).}
  \label{fig:spec}
\end{figure}

\begin{figure}[tp]
  \centering
  {\includegraphics[width=.97\linewidth,height=.2\linewidth]{%
      growth_grerm_bar}}\\[-.08\linewidth]
  {\includegraphics[width=\linewidth]{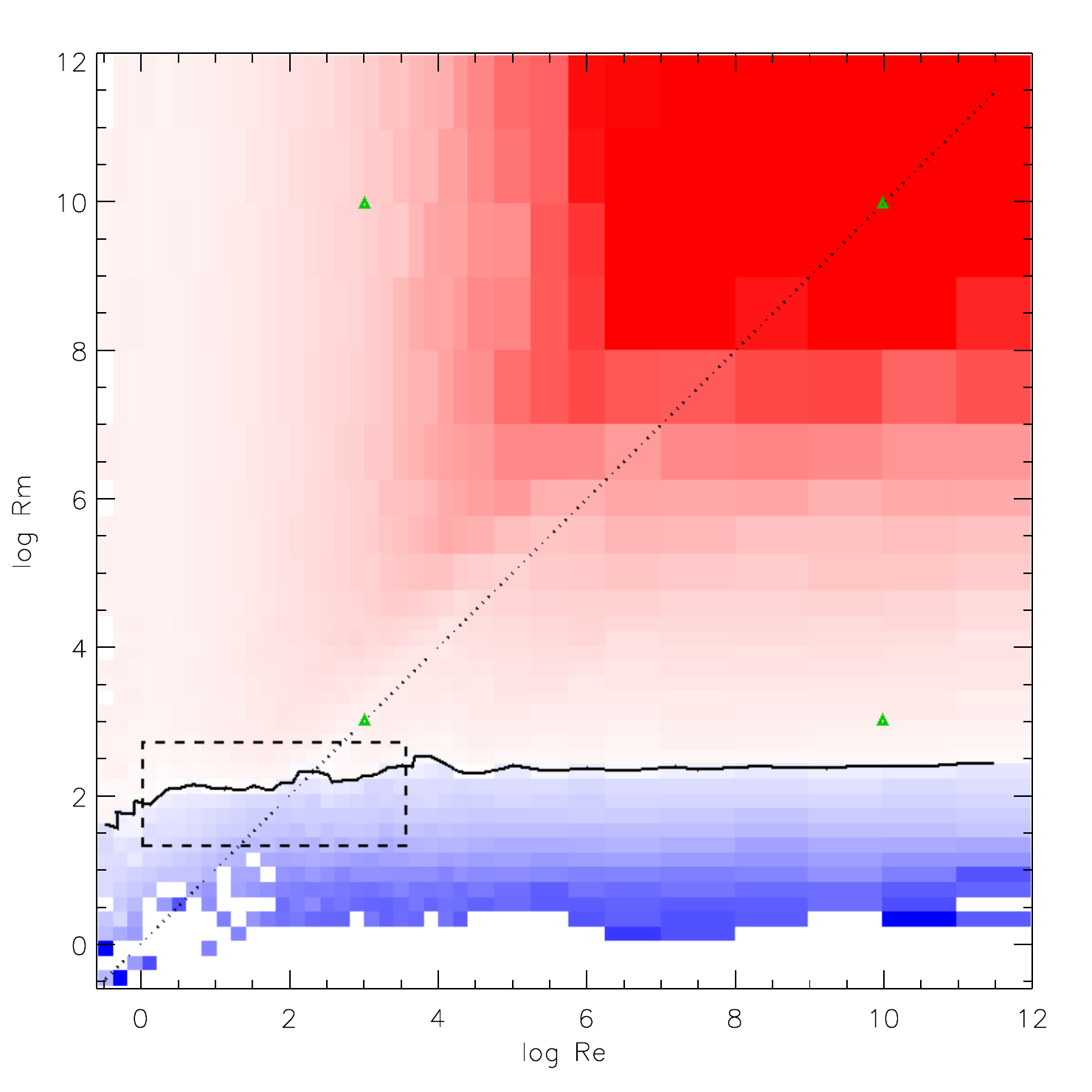}}
  \caption{(online only) Same as Fig.~\ref{fig:gamma} for the local GOY model.}
  \label{fig:gamma2}
\end{figure}

\begin{figure}[tp]
  \centering
  {\includegraphics[width=.97\linewidth,height=.2\linewidth]{%
      growth_grerm_bar}}\\[-.08\linewidth]
  {\includegraphics[width=\linewidth]{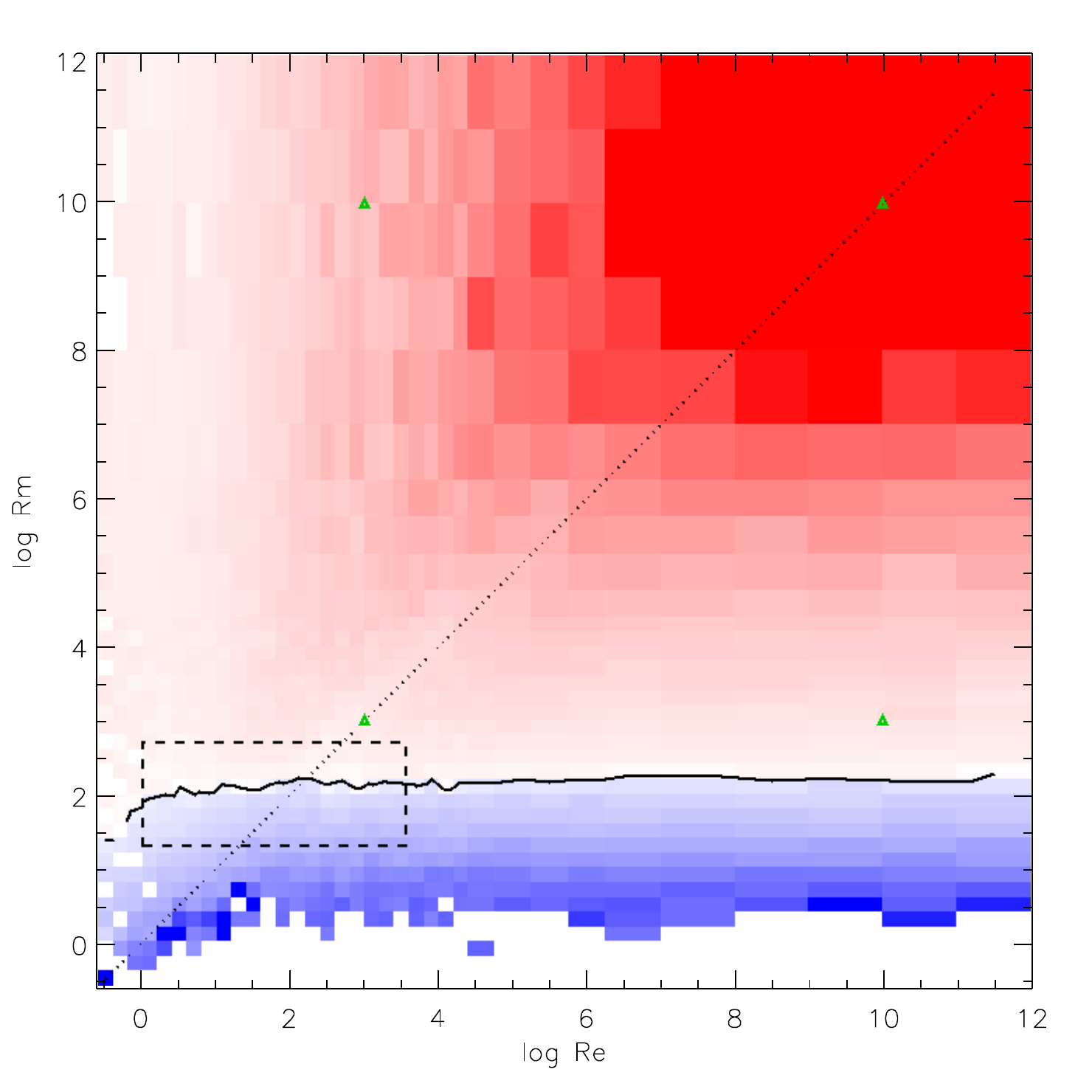}}
  \caption{(online only) Same as Fig.~\ref{fig:gamma} for the strongly nonlocal ($\alpha=-1$) Sabra model.}
  \label{fig:gamma3}
\end{figure}

The difference, however small, between local and nonlocal models is surprising if one considers that nonlocal interactions in wavenumber space are expected to become important at large \prn.  However, for all values of $\prn$, the evolution of magnetic spectra is consistent with mainly local transfers from kinetic to magnetic energy, and so does not require important nonlocal transfers:
\begin{enumerate}
  \item For $\prn = 1$ ($\nu = \eta = 10^{-3}$ or $10^{-10}$ in Fig.~\ref{fig:spec}), magnetic energy grows first on scales corresponding to the end of the inertial range of the kinetic energy spectrum, close to the dissipation range, at a speed depending on Reynolds numbers.

  \item For $\prn \ll 1$ ($\nu = 10^{-10}$ and $\eta = 10^{-3}$ in Fig.~\ref{fig:spec}), the wavenumber range where magnetic energy can grow is limited by the magnetic diffusivity scale.

  \item For $\prn \gg 1$ ($\nu = 10^{-3}$ and $\eta = 10^{-10}$ in Fig.~\ref{fig:spec}), the range where magnetic energy can grow is limited by the viscous scale, showing that transfers from the kinetic to the magnetic energy are still mainly local (despite the inclusion of nonlocal terms in the model).
\end{enumerate}
We note that the scales on which the magnetic energy grows most are consistent, at both small and large $\prn$, with the scales used above to compute the magnetic energy growth rate in the small-scale dynamo phenomenology.

\section{Conclusion}
\label{sec:conc}

We have computed the growth rate $\gamma$ of the magnetic field in the kinematic regime of a dynamo as a function of the Reynolds numbers $\ren$ and $\rmn$, with, thanks to shell models, a much wider range of parameters than previous studies using direct numerical simulations.  This wider parameter range is of astrophysical relevance and brings a new perspective to results from direct numerical simulations; it allows us to answer some important outstanding questions about the kinematic regime of the dynamo, assuming of course that results from shell models remain valid for the general MHD equations.

We confirm that the critical magnetic Reynolds number $\rmn_c$ tends to a finite value at large Reynolds numbers $\ren$.  Furthermore, the growth rate $\gamma$ tends to a finite value $l(\rmn)$ when $\ren$ tends to infinity, and we find a scaling $l(\rmn) \sim \rmn^{\beta}$ with $\beta\approx 0.4$.  Similarly, $\gamma$ tends to a finite value $l'(\ren) \sim \ren^{\beta'}$ when $\rmn$ tends to infinity, with $\beta'\approx 0.4$.  Both limits $l(\rmn)$ and $l'(\ren)$ are attained for $\rmn \gtrsim \ren$.

These scalings can be explained by an intermittent, mainly small-scale dynamo. Furthermore, our results imply that nonlocal interactions (in Fourier space) play a role in the kinematic dynamo, although a limited one. This behavior is consistent with results on the locality of nonlinear interactions in MHD systems: in \citet{alexakis05}, \citet{mininni05} (direct numerical simulations), and \citet{plunian07} (shell models), nonlocal interactions are significant but are mostly confined to a wavenumber range that is relatively small compared to the wavenumber separation factor between shells that interact in shell models.

\begin{acknowledgements}
  Financial support from STFC (UK), CNES and CNRS (France) is acknowledged. I thank the referee for his useful comments. This work has benefited from discussions with P.\ Démoulin, S.\ Fauve,  S.\ Galtier, F.\ Plunian, M.\ Proctor, and A.\ Schekochihin,  in particular at the 2009 workshop ``Frontiers in Dynamo Theory'' in Paris, and from the work of K.\ Olivier during an internship in Orsay.
\end{acknowledgements}

\bibliographystyle{aa}
\bibliography{solphys,books}

\clearpage

\end{document}